\documentclass{article}

\usepackage[english]{babel}

\usepackage[letterpaper,top=2cm,bottom=2cm,left=3cm,right=3cm,marginparwidth=1.75cm]{geometry}

\usepackage{amsmath}
\usepackage{listings}
\usepackage{graphicx}
\usepackage{float}
\usepackage[colorlinks=true, allcolors=blue]{hyperref}

\title{PulseNet: Deep Learning ecg-signal classification using random augmentation policy and continous wavelet transform for canines}
\author{Andre Dourson$^1$, Roberto Santilli$^{3,5}$, Federica Marchesotti$^3$, Jennifer Schneiderman$^4$, 
\\   Oliver Roman Stiel$^2$, Fernando Junior$^1$, Michael Fitzke$^1$, Norbert Sithirangathan$^1$,
\\   Emil Walleser$^1$, Xiaoli Qiao$^1$, Mark Parkinson$^1$\\ }

\date{%
    $^1$Mars Digital Technologies\\%
    $^2$ Mars Petcare\\%
    $^3$Anicura\\%
    $^4$Antech Imaging Services\\%
    $^5$Cornell University\\[2ex]%
    May 16, 2023
}

\usepackage{fancyhdr}
\graphicspath{ {./images/} }
\usepackage{graphicx}
\usepackage{amsmath}
\usepackage{amssymb}
\usepackage{lscape} 
\usepackage{geometry}
\usepackage{mathrsfs}
\usepackage{subfig}
\usepackage{mdframed}
\usepackage{float}
\usepackage{makecell}
\usepackage{adjustbox}
\usepackage{tabularx}
\usepackage[numbers]{natbib}  
\usepackage{hyperref}  

\begin{document}
\maketitle

\begin{abstract}
\par
Evaluating canine electrocardiograms (ECG) require skilled veterinarians, but current availability of veterinary cardiologists for ECG interpretation and diagnostic support is limited. Developing tools for automated assessment of ECG sequences can improve veterinary care by providing clinicians real-time results and decision support tools. We implement a deep convolutional neural network (CNN) approach for classifying canine electrocardiogram sequences as either normal or abnormal. ECG records are converted into 8 second Lead II sequences and classified as either normal (no evidence of cardiac abnormalities) or abnormal (presence of one or more cardiac abnormalities). For training ECG sequences are randomly augmented using RandomAugmentECG, a new augmentation library implemented specifically for this project. Each chunk is then is converted using a continuous wavelet transform into a 2D scalogram. The 2D scalogram are then classified as either normal or abnormal by a binary CNN classifier. Experimental results are validated against three boarded veterinary cardiologists achieving an AUC-ROC score of 0.9506 on test dataset matching human level performance. Additionally, we describe model deployment to Microsoft Azure using an MLOps approach. To our knowledge, this work is one of the first attempts to implement a deep learning model to automatically classify ECG sequences for canines.Implementing automated ECG classification will enhance veterinary care through improved diagnostic performance and increased clinic efficiency.

\end{abstract}
\section{Introduction}
\par
The role of electrocardiography (ECG) in the veterinary diagnostic work-up had increased in the last decades becoming an important tool to diagnose different primary and secondary rhythm disturbances such as atrial fibrillation, focal atrial tachycardias, junctional tachycardias, reciprocating tachycardias, atrial flutter, and ventricular arrhythmias and conduction abnormalities. \cite{jvc}\cite{jvim}\cite{jvim2}\cite{jvim3}\cite{jvc2}\cite{jvc31}\cite{jvc4}\cite{jvc5}\cite{jvc6}\cite{jsap}\cite{vj} \\ \cite{jvc7}\cite{jvc8} The analysis of ECG appearance is complex and diagnosis of specific arrhythmias often requires veterinary cardiologist review. However, the shortage of trained veterinary specialists makes this challenging.\cite{wogan_2022}

\par
Electrocardiogram (ECG) shows cardiac electrical activities, named vectors, projected on the body surface allowing the analysis of cardiac arrhythmias, ischemia, and conduction abnormalities.\cite{electrocardiography}
\par

The cardiac vectors can be analyzed on the frontal plane using the 6 lead system (I,II,III, aVR, aVL, aVF) and on the horizontal plane using the precordial system (V1,V2,V3,V4, V5, V6) modified for the dog. \cite{electrocardiography} \cite{amjres}. They appear as consecutive waves (atrial – P wave and ventricular – QRS complex) displayed on a millimeter paper at different programmed speed and calibration.\cite{electrocardiography}
\par

\par
For this reason, with the help of artificial intelligence (AI) the interpretation process can be speed up, differentiating first normal from abnormal electrocardiogram and then grouping them in several subgroups of arrhythmias then submitted to the cardiologists for a final diagnosis. 

\section{Related Work}
The problem of automated ECG classification is well-studied in the field of human medicine, and numerous deep learning models have been proposed to tackle this task (\cite{acharya2018deep}\cite{hannun2019cardiologist} \cite{yildirim2018arrhythmia}). However, the problem of automated ECG classification for canines has received limited attention in the literature, and few deep learning models have been proposed for this specific purpose. Previous work with automated classification of canine ECG signals has focused on extracting ECG waveform details and applying a decision algorithm for categorizing abnormalities \cite{Estrada2021-xk}. 

The current state-of-the-art in automated ECG classification for humans often relies on deep convolutional neural networks (CNNs) and various data augmentation techniques (\cite{kachuee2018ecg}, \cite{clifford2017af}). Our work builds upon these approaches by adapting them for canine ECG signal classification.

Our work shares similarities with the recent development of ECG-based deep learning models for other non-human animals, such as horses (\cite{van2020transfer}) and mice (\cite{liao2023deepmicetl}). These studies demonstrate the potential of leveraging deep learning approaches for animal ECG classification. Our study contributes to this emerging field by focusing on canines and exploring the impact of random augmentation policies and continuous wavelet transforms on classification performance.

Zero-shot transfer from humans to canines are discussed in \cite{thomas2021ecg} with results indicating that dedicated models trained on canine are needed to ensure reliable predictions. This work fills that gap by presenting a novel contribution to the field of automated ECG classification for canines. By incorporating state-of-the-art techniques from human ECG analysis and introducing new methods specifically tailored for canine ECG signals, we demonstrate the potential of deep learning models to improve diagnostic performance and clinical efficiency in veterinary medicine.

\section{Data}

\subsection{DICOM files}

The ECG records used in this project were collected from 1462 distinct DICOM files, containing 6 or 12 leads and all sampled at 500Hz. For this project we utilized only lead II, which has shown comparable performance to inclusion of additional leads \cite{physionet21}.

The collected ECG raw sequences have various durations (Figure 1) :
\begin{figure}[H]
\centering
\includegraphics[width=0.7\textwidth]{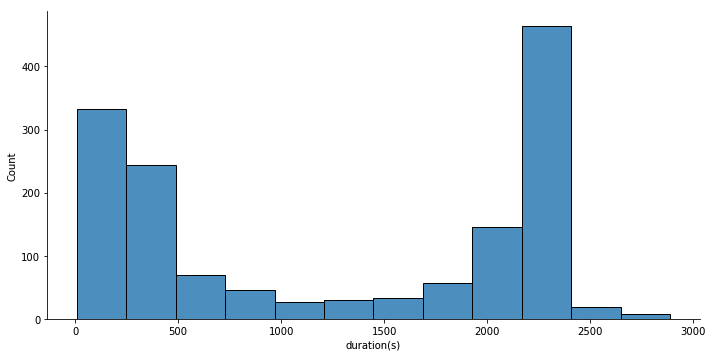}
\caption{Distribution of ECG sequence length (in seconds)}
\label{fig:durations}
\end{figure}

In order to prepare our dataset for training, the ecg signals were extracted from DICOM format into one dimensional arrays. The extracted signals were then pre-processed using the following pipeline (Figure 2) :

\begin{figure}[H]
\centering
\includegraphics[scale=0.3]{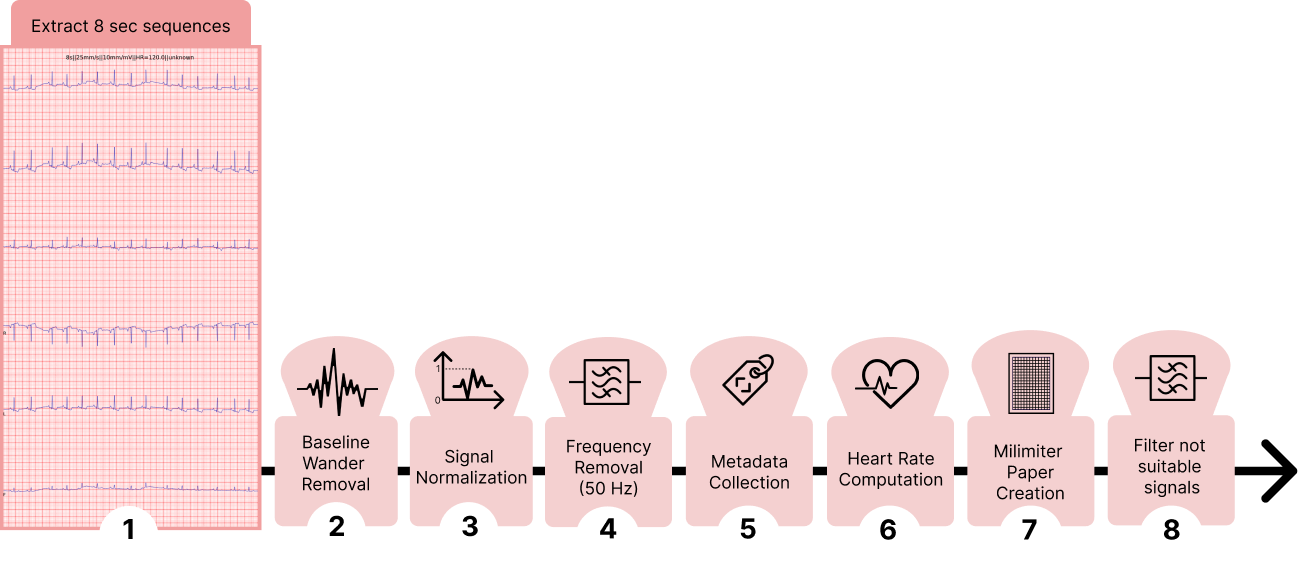}
\caption{Pre-processing steps}
\label{fig:preprocessing_pipeline}
\end{figure}

\begin{itemize}
\item (1) Extraction of 8 seconds sequences with an overlap of 1 second between each sequence
\item (2) Baseline wander removal
\item (3) Signal normalization between 0 and 1
\item (4) Frequency removal (50Hz)
\item (5) Collection of metadata (race, breed, age, weight etc...)
\item (6) Automatic computation of the heart rate
\item (7) Creation of (millimeter paper look and feel) 8 seconds snippets for experts review
\item (8) Removal of signals which are not suitable for diagnostics
\end{itemize}

\begin{figure}[H]
\includegraphics[width=0.7\textwidth]{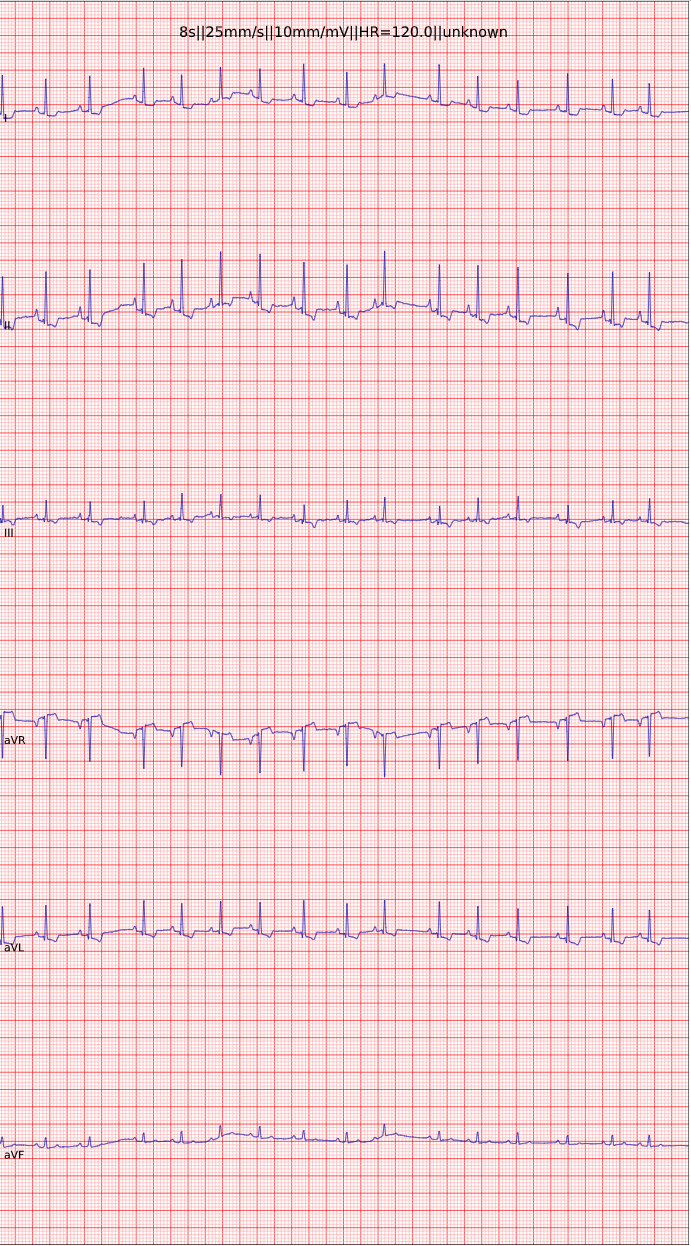}
\centering
\caption{Example of 8 seconds snippet}
\label{fig:ECG}
\end{figure}
We extracted a total of 222,847 sequences of 8 seconds (Lead II signal) following the previously described steps (Figure 3). These 8 seconds signal chunks were then categorized into seven categories (normal, unsuitable for diagnosis, supraventricular arrhythmias, ventricular arrhythmias, bradyarrhythmias, and conduction disorders by veterinary board certified cardiologists.(Figure 4).Unsuitable for diagnostic chunks were removed from the dataset, and all arrhythmias were then grouped into a single abnormal category.

\begin{figure}[H]
\includegraphics[width=0.7\textwidth]{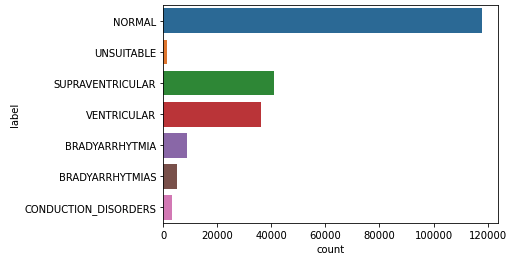}
\centering
\caption{ECG sequences labels}
\label{fig:sequences}
\end{figure}

\subsection{RandAugmentECG library}

Data augmentation of image data is a key strategy to improve neural network performance \cite{Shorten2019}. Similar performance improvements may be possible for one-dimensional data, like ECG segments. In this paper we introduce RandAugmentECG, a PyTorch ECG augmentation library inspired from the work done by \cite{rand} and \cite{monai} and a one-dimensional equivalent to the library developed for images \cite{randaug}.

This PyTorch library provides custom augmentation listed below :

\begin{itemize}
\item RandShift1d : randomly shifts a portion of the signal
\item RandScale1d : randomly scales the signal by a given factor
\item RandRoll1d : rolls one random portion of the signal
\item RandDrop1d : randomly drops one consecutive portion of the signal, replacing with a specific number (zero for instance)
\item RandAddSine1d : adds a randomly generated sinusoidal signal to the source signal
\item RandAddSinePartial1d : adds a randomly generated sinusoidal signal to a random consecutive portion of the source signal
\item RandAddSquarePulse1d : adds a random pulse square signal to the source signal
\item RandAddSquarePulsePartial1d : adds a random pulse square signal to a random consecutive portion of the source signal
\item RandAddGaussianNoise1d : adds a random gaussian noise to the source signal
\item ToTensor1d : converts source signal to 1D tensor
\item Resample1d : resamples the source signal to given frequency
\item Normalize1d : normalizes the source signal between 0 and 1
\item Standardize1d : standardizes the source signal
\item ZeroPad1d: if the signal is shorter than the expected length, pads the signal with zeros
\item FillEmpty1d : if the signal contains NaN elements, replaces these elements with zeros
\item RemoveBaselineWander1d : removes signal wander line
\item RemoveFrequency1d : removes a given frequency from the signal
\item ToCWT : converts the 1D signal to a 2D scalogram (continuous wavelet transform)
\item ToSTFT : converts the 1D signal to a 2D spectrogram (Short time Fourier Transform)
\item ToCWFT : converts the 1D signal a mix of CWT and STFT (either summing up or concatenating)
\end{itemize}

The purpose of RandAugmentECG is similar to the work done by for 1D signal and for 2D images. It consists on applying a stochastic combination of several 1 dimensional transforms present in the library :

\begin{itemize}
\item RandShift1d
\item RandScale1d
\item RandRoll1d
\item RandDrop1d
\item RandAddSine1d
\item RandAddSquarePulse1d
\item RandAddGaussianNoise1d
\end{itemize}

The maximum number of individual transforms to be applied along with the magnitude of the change are governed by two parameters \textit{n} and \textit{m} (Figure 5)

\begin{figure}[H]
\includegraphics[scale=0.16]{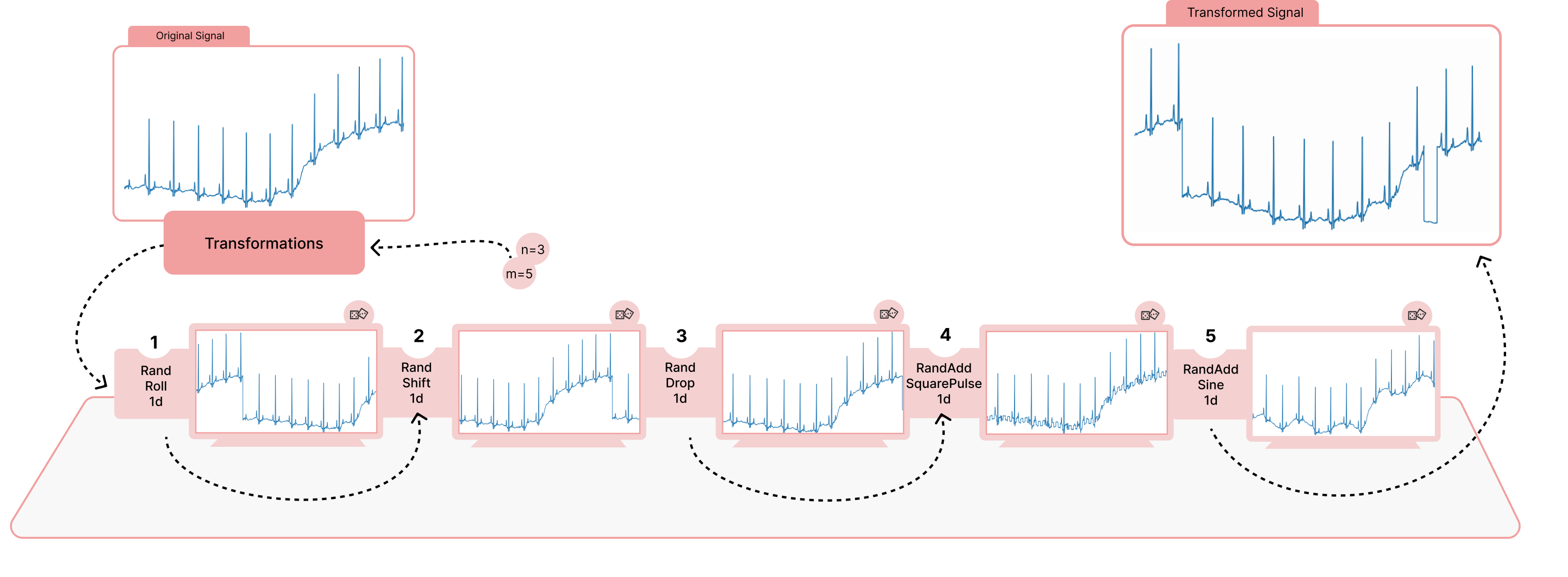}
\centering
\caption{RandAugmentECG samples (n=3,m=5)}
\label{fig:randaugment}
\end{figure}

\subsection{Continuous Wavelet transform}

Wavelet transform\cite{wavelet}\cite{wavelet1} and Short Time Fourier transform\cite{fft} are widely used in signal processing applications. As opposed to Fourier transform where we analyse the time/frequency domain for a window (portion) of the signal and then do a sliding of this window to reiterate the same analysis before eventually recombining all resulting transforms, Continuous Wavelet transform has the benefit to be able to process the entirety of the signal at once, modulating the wavelet dilation and location, and doing a convolution between the resulting wavelet function and the signal (Figure 6).

\begin{figure}[H]
\includegraphics[width=1.0\textwidth]{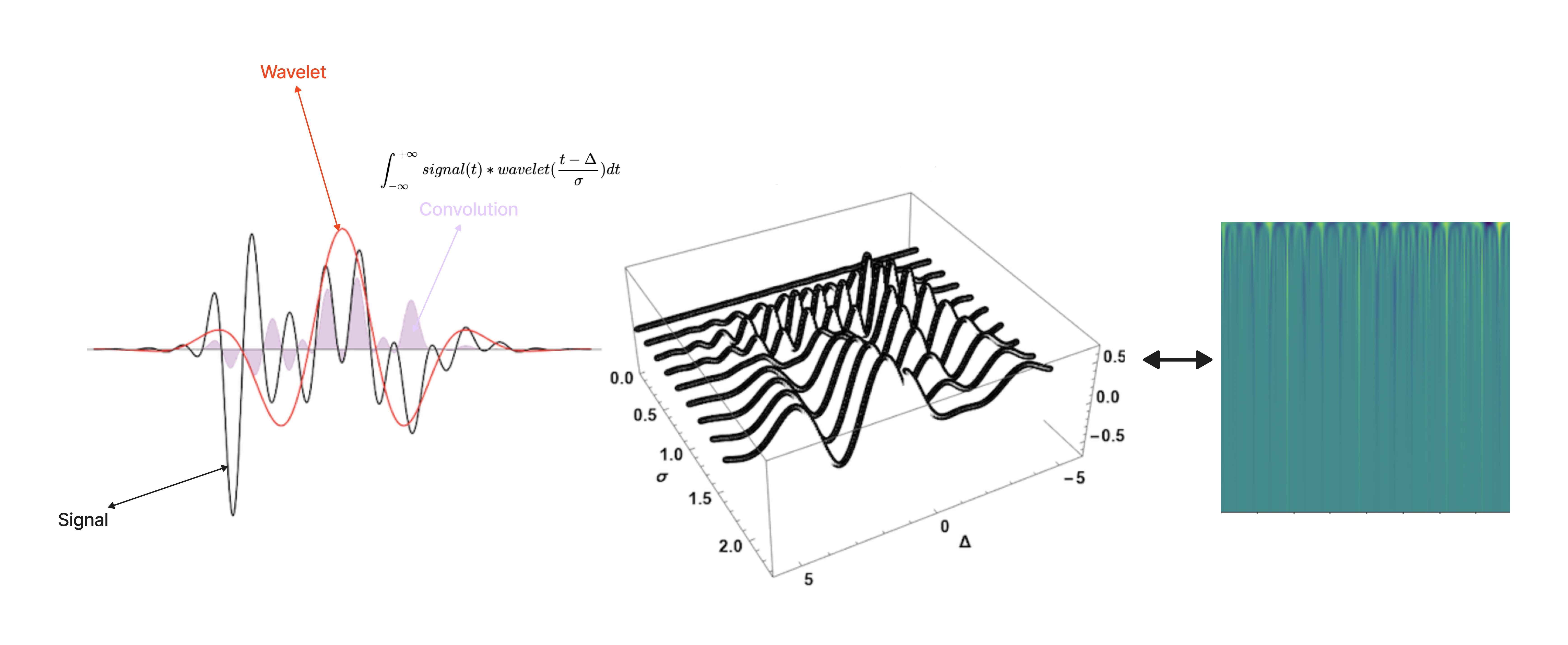}
\centering
\caption{convolution of wavelets with input signal to render the scalogram}
\label{fig:scalogram}
\end{figure}

If x(t) is a continuous signal, at a scale $\sigma>0$ , $\sigma\in\mathbb{R^{+*}}, \Delta\in\mathbb{R}$
then its wavelet transform is defined as

\begin{equation*}
    X_w(\sigma,\Delta) = \frac{1}{|\sigma|^{1/2}} \int_{-\infty}^{+\infty} x(t)\overline{\psi}(\frac{t-\Delta}{\sigma})dt
\end{equation*}

where $\psi(t)$ is a continuous function in both the time domain and the frequency domain called the mother wavelet, and $\overline\psi$ is its complex conjugate. 
$\sigma$ is also called the dilation parameter of the wavelet and $\Delta$ is called the location parameter of the wavelet

The purpose of the mother wavelet is to provide a new version of it called the daughter wavelet, which is a translated and scaled version of the mother wavelet (defined by the $\sigma$ and $\delta$ parameters). 

For ECG, a particular type of wavelet seems intuitively to be appropriate (as it somewhat mimics the ECG shape): the Ricker wavelet, a.k.a. Mexican hat wavelet (due to its shape of a sombrero)
The Mexican hat wavelet (Figure 7) is defined as the second derivative of a Gaussian function given by 

\begin{equation*}
    \psi(t) = \frac{2}{\sqrt{3\sigma}\pi^4}(1-(\frac{t}{\sigma})^2)e^-\frac{t}{2\sigma^2}
\end{equation*}

\begin{figure}[H]
\includegraphics[width=.5\textwidth]{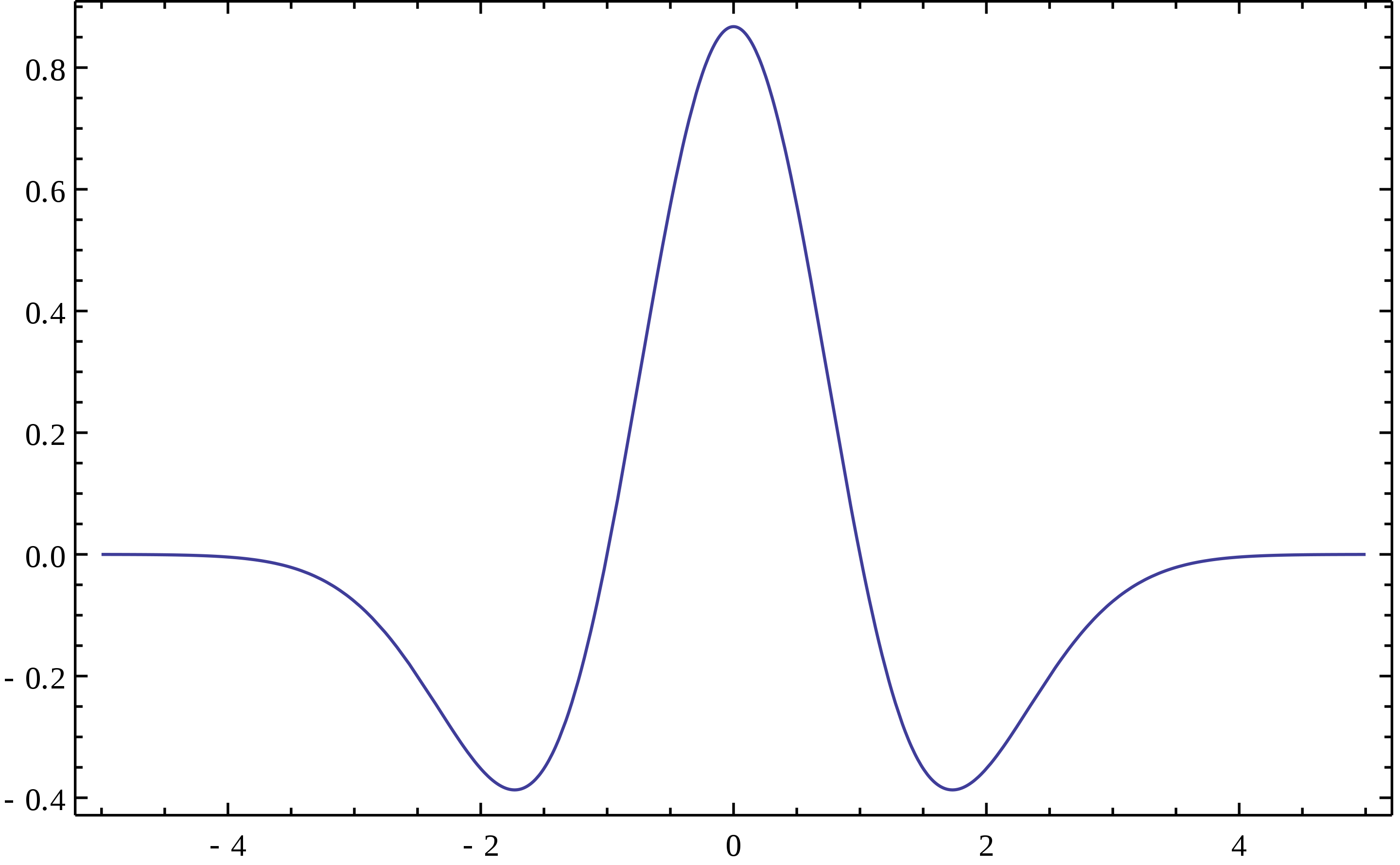}
\centering
\caption{Ricker wavelet a.k.a. Mexican hat wavelet}
\label{fig:mexican}
\end{figure}

\subsection{Creation of training dataset}

We split the 222,847 ECG records into a training and a validation dataset (99\%/1\%). On the training set, we first applied the Resample1d transform to resample each signal to 250Hz in order to reduce the computation cost. Next, we further transformed the signals by stochastically augmenting each signal using the RandAugmentECG transform (n=5, m=5). We finally converted the augmented signals using the ToCWT transform into a 2D scalogram of size 300x300 (Figure 8). 

On the validation set, we only applied the Resample1d transform to resample signals to 250Hz to remain consistent with our training dataset.

Using this approach we are in turn converting this task into a more standard image classification problem. 

\begin{figure}[H]
\includegraphics[width=0.8\textwidth]{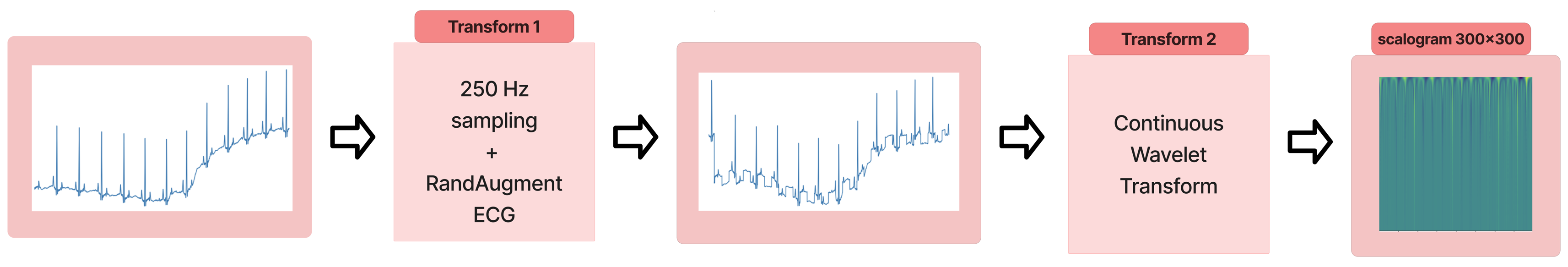}
\centering
\caption{scalogram creation}
\label{fig:ScalogramCreation}
\end{figure}

\section{Models and training details}

\subsection{Model Architecture}

\par We initially trained our model using standard types of convolutional neural networks architecture (ResNet18, ResNet34, ResNet50) but we also designed our own architectures. One custom architecture, named "R" model (Figure 10 and Table 2), was inspired by the ResNet architecture \cite{resnet} but with only 3 ResNet blocks, and with the replacement of Max Pooling with Average Pooling\cite{pooling}. The second architecture, designated type "M" (Figure 9 and Table 3) is composed of 3 layers, each consisting of a 2D convolutional layer followed by Batch Normalization, ReLU activation and Average pooling. Both "M" and "R" models are implemented with Dropout in the final classification layer (p = 0.2). From our experiments, the best performance as determined by area under the receiver operating characteristic curve (AUC)  were achieved using the "M" architecture, followed closely by "R" architecture.
\par The intuition behind replacement of Max Pooling with Average Pooling in our models could be that Average Pooling would be more likely be to extract useful features from scalograms as transition/edges are less abrupt in scalograms vs regular images (where contours are maybe more apparent).
(Table 1)

\begin{table}[H]
\centering
\begin{tabular}{ |p{2.5cm}||p{2.5cm}| }
 \hline
 \multicolumn{2}{|c|}{Max AUC per model architecture} \\
 \hline
  Architecture & Max AUC\\
 \hline
  Resnet18 & 0.906\\
  Resnet34 & 0.922\\
  Resnet50 & 0.922\\
  "R" model & \textbf{0.941}\\
  "M" model & \textbf{0.951}\\
 \hline
\end{tabular}
\caption{Models architecture : comparison of AUC value}
\end{table}

\begin{figure}[H]
\includegraphics[width=.9\textwidth]{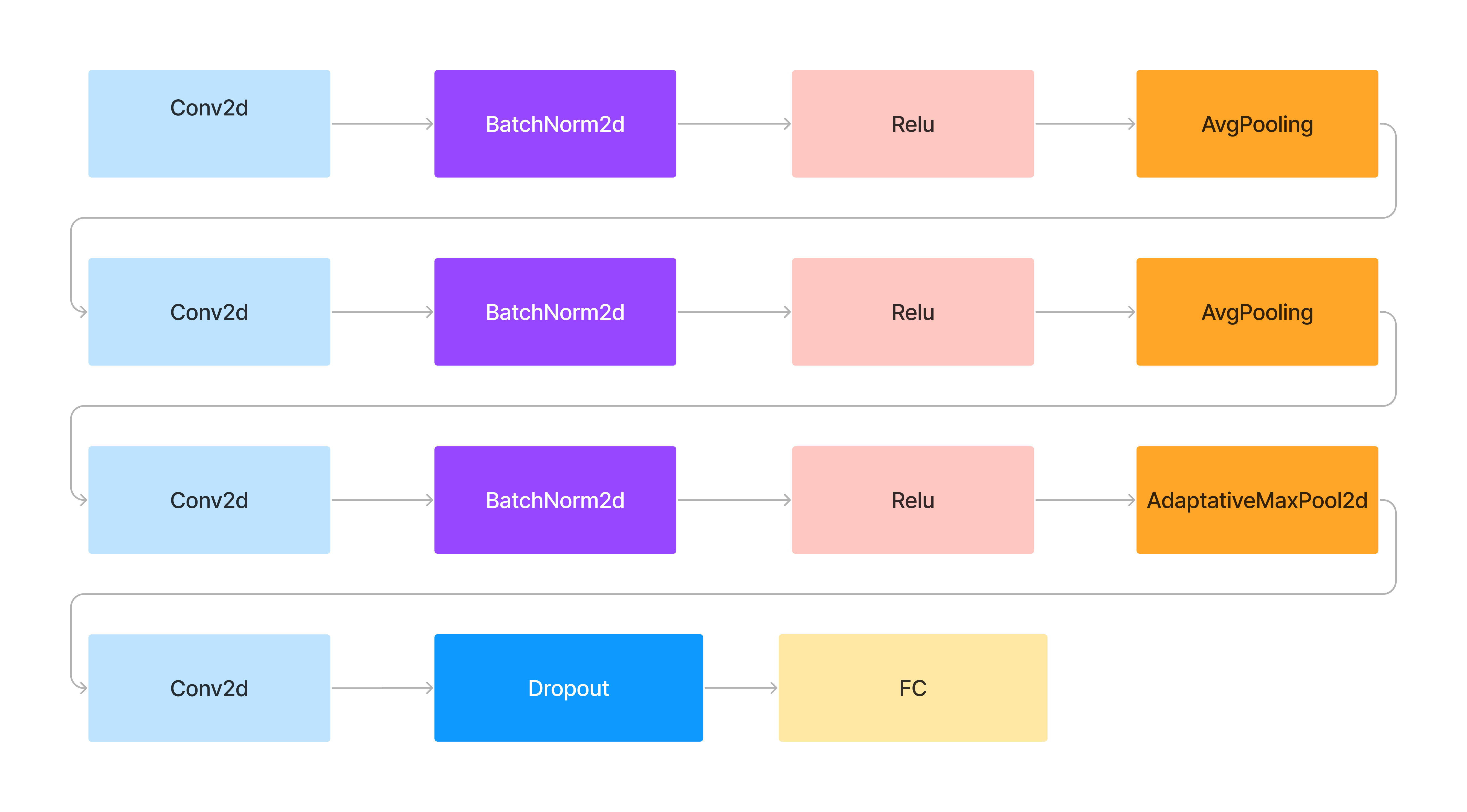}
\centering
\caption{"M" model}
\label{fig:MModel}
\end{figure}

\begin{figure}[H]
\includegraphics[width=.9\textwidth]{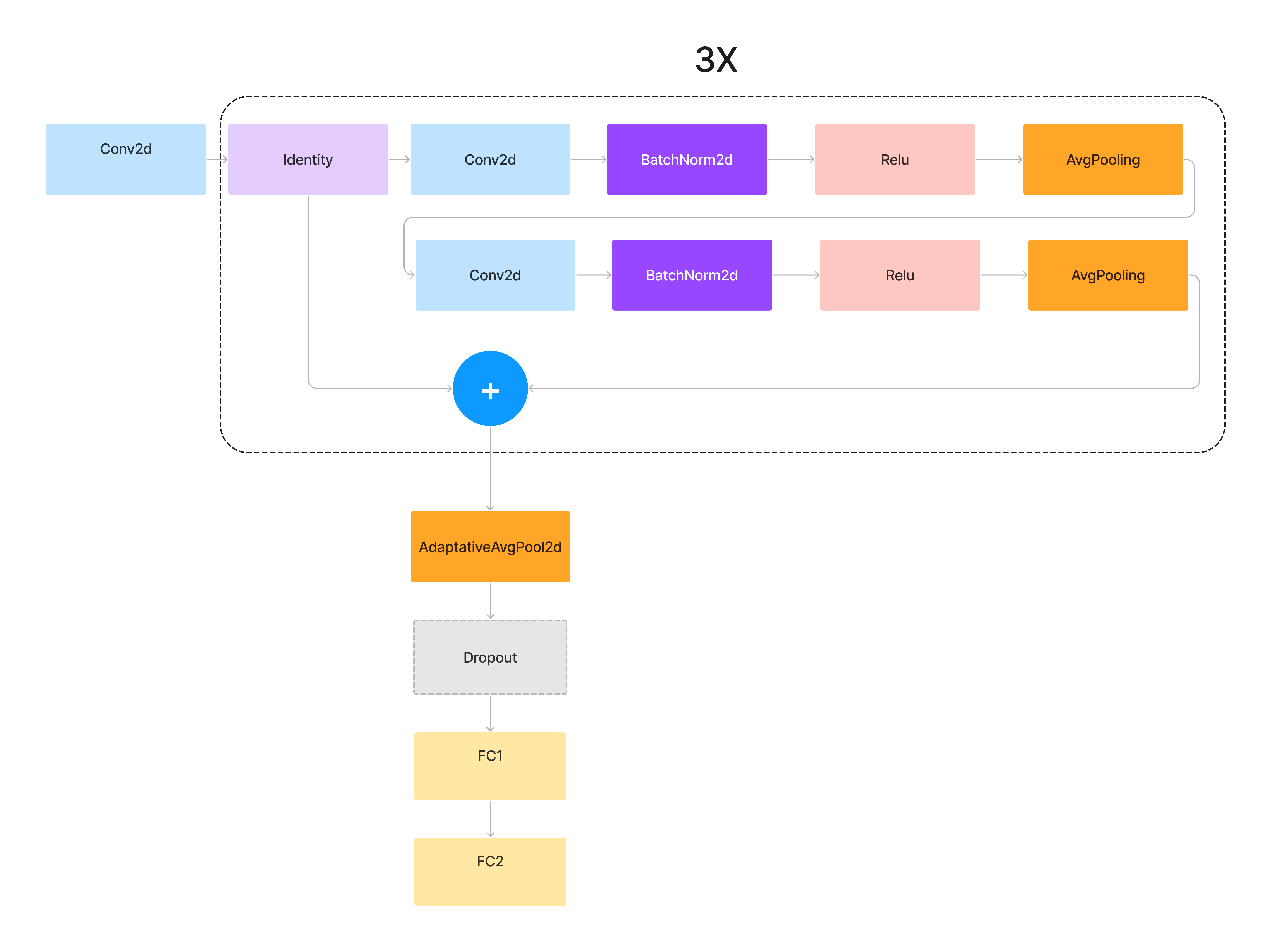}
\centering
\caption{"R" model}
\label{fig:RModel}
\end{figure}

\begin{table}[H]
\begin{adjustbox}{width=\textwidth,center}
\begin{tabularx}{1.0\linewidth}{ |X||X|X|X|X|X|X|  }
 \hline
 \multicolumn{7}{|c|}{Architecture of "R" Model} \\
 \hline
 Layer& Type &Kernel size&Stride&Padding&Features&Input size\\
 \hline
 Layer1 & Conv2d  &7 x 7&  2 &3& 64 & 1 x 300 x 300\\
 Layer2 & Conv2d  &3 x 3&  1 & 1& 32 & 1 x 300 x 300\\
 Layer3 & Conv2d  &3 x 3&  1 & 1& 64 & 32 x 300 x 300\\
 Layer4 & AvgPool2d  &2 x 2& 2  &0 & - & 64 x 300 x 300\\
 Layer5 & Conv2d  &3 x 3&  1 & 1 & 128 & 64 x 150 x 150\\
 Layer6 & Conv2d  &3 x 3&  1 & 1 & 256 & 128 x 150 x 150\\
 Layer7 & AvgPool2d  &2 x 2&  2 & 0 & - & 256 x 152 x 152\\
 Layer8 & Conv2d  &3 x 3&  1 & 1& 512 & 256 x 76 x 76\\
 Layer9 & Conv2d  &3 x 3&  1 &1 & 1024 & 512 x 76 x 76\\
 Layer10 & AvgPool2d  &2 x 2&  2 & 0 & - & 1024 x 76 x 76\\
 Layer11 & AdpAvP2d  & - &  -  & -  & 1024 & 1024 x 38 x 38\\
 Layer12 & Linear & & & &256 & 1024 x 1 x 1\\
 Layer13 & Linear & & &  & 2 & 256\\
 \hline
\end{tabularx}
\end{adjustbox}
\caption{Architecture of "R" model}
\end{table}

\begin{table}[H]
\begin{adjustbox}{width=\textwidth,center}
\begin{tabularx}{1.0\linewidth}{ |X||X|X|X|X|X|X|  }
 \hline
 \multicolumn{7}{|c|}{Architecture of "M" Model} \\
 \hline
 Layer& Type &Kernel size&Stride&Padding&Features&Input size\\
 \hline
 Layer1   & Conv2d  &7 x 7&  1 & 0 &128 & 1 x 300 x 300\\
 Layer2 & AvgPool2d  &5 x 5&  2 & 0 & - & 128 x 294 x 294\\
 Layer3   & Conv2d  &3 x 3&  1 & 0 & 256 & 128 x 58 x 58\\
 Layer4   & AvgPool2d  &3 x 3&  2 & 0 & - & 256 x 56 x 56\\
 Layer5   & Conv2d  &3 x 3&  1 & 0 & 512 & 256 x 18 x 18\\
 Layer6   & AdMaxP2d  & - &  - & - & 512 & 512 x 16 x 16\\
 Layer7   & Conv2d  &1 x 1&  1 & 0 & 512 & 512 x 1 x 1\\
 Layer8   & Linear  & - &  - & - & 2 & 512\\
 \hline
\end{tabularx}
\end{adjustbox}
\caption{Architecture of "M" model}
\end{table}

\subsection{Training details}

In all our experiments, we used PyTorch Lightning \cite{ptl} to train and evaluate our models. The 2 models were each trained using a Tesla V100 system (16 Gb). We trained using mixed precision for 100 epochs with a batch size of 110 (for "M" model) and 80 (for "R" model), and we applied a gradient accumulation policy as optimization occurred only every 3 mini batches. Both models were initialized using Xavier initialization \cite{xavier} and a Softmax activation was applied on the final layer. We used an Adam optimizer with a initial learning rate of 3e-4 with a weight decay of 0.2. The learning rate was decayed using CosineAnnealing with warm restart (Figure 11), with the maximum learning rate of 3e-4 and a $\tau^0$ of 5000.

\begin{figure}[H]
    \centering
    \includegraphics[width=.5\textwidth]{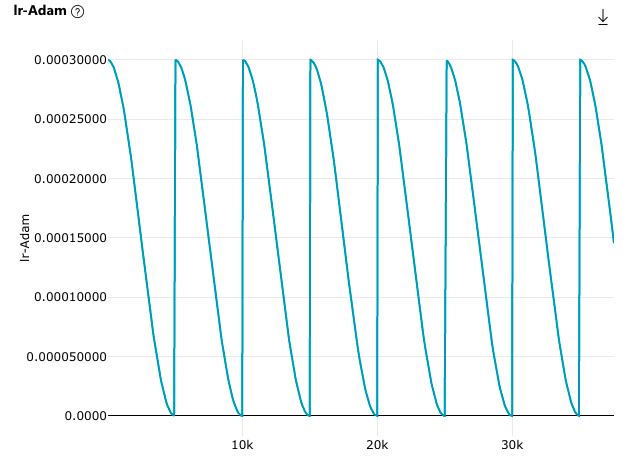}
    \caption{Cosine Annealing with warm restart}
    \label{fig:my_label}
\end{figure}

We trained the models using CrossEntropyLoss\cite{crossentropy}, but also with the PolyLoss described in\cite{polyloss}. Our experiments show the best results with PolyLoss (epsilon=2.5 or 3.5).

\begin{equation*}
    PolyLoss = -\sum_{n=1}^{k} y_i*log(\hat{y_i}) + \epsilon * (1-\hat{y_i})
\end{equation*}

\section{Testing}

All testing was performed on a "gold" dataset containing 808 sequences of 8 seconds extracted from different DICOM files than the ones used to train and validate the model. These sequences were labeled independently by three board certified veterinary cardiologists. Gold standard or ground truth was determined using majority vote between the three cardiologist assessment (normal vs. abnormal). We benchmarked our best performing model (determined by AUC on our validation dataset), against human cardiologist performance.
We then used our best model to benchmark it against the human expert using the ROC curve, using a threshold computed using Youden methodology to determine the ideal cutoff to distinguish between normal and abnormal classes. Our best model ("M" model) achieved an AUC score of 0.9506 and a F1 score (weighted) of 89,28\%. Precision is 86,28\% , Recall is 92,5\%, Accuracy is 91,95\%

Comparing with the cardiologists, the model is on par with 2 of them with respect to the F1 score, and doing better than 2 of them in term of Recall and Accuracy (Table 4, Figure 12 and Figure 13)

\begin{table}[H]
\centering
\begin{tabular}{ |p{2cm}||p{2cm}||p{2cm}||p{2cm}||p{2cm}|}
 \hline
 \multicolumn{5}{|c|}{Comparison model vs cardiologists} \\
 \hline
 Metric & Model & Cardiologist0 & Cardiologist1 & Cardiologist2\\
 \hline
 F1 & \textbf{89,28\%} & 88,37\% & 97,89\% & 89,85\% \\
 Precision & \textbf{86,28\%} & 98,46\% & 97,51\% & 95,11\% \\
 Recall & \textbf{92,50\%} & 80,15\% & 98,28\% & 85,15\% \\
 Accuracy & \textbf{91,95\%} & 83,63\% & 96,72\% & 85,09\%\\
 \hline
\end{tabular}
\caption{Best Model vs Cardiologists}
\end{table}

\begin{figure}[H]
\includegraphics[width=1.0\textwidth]{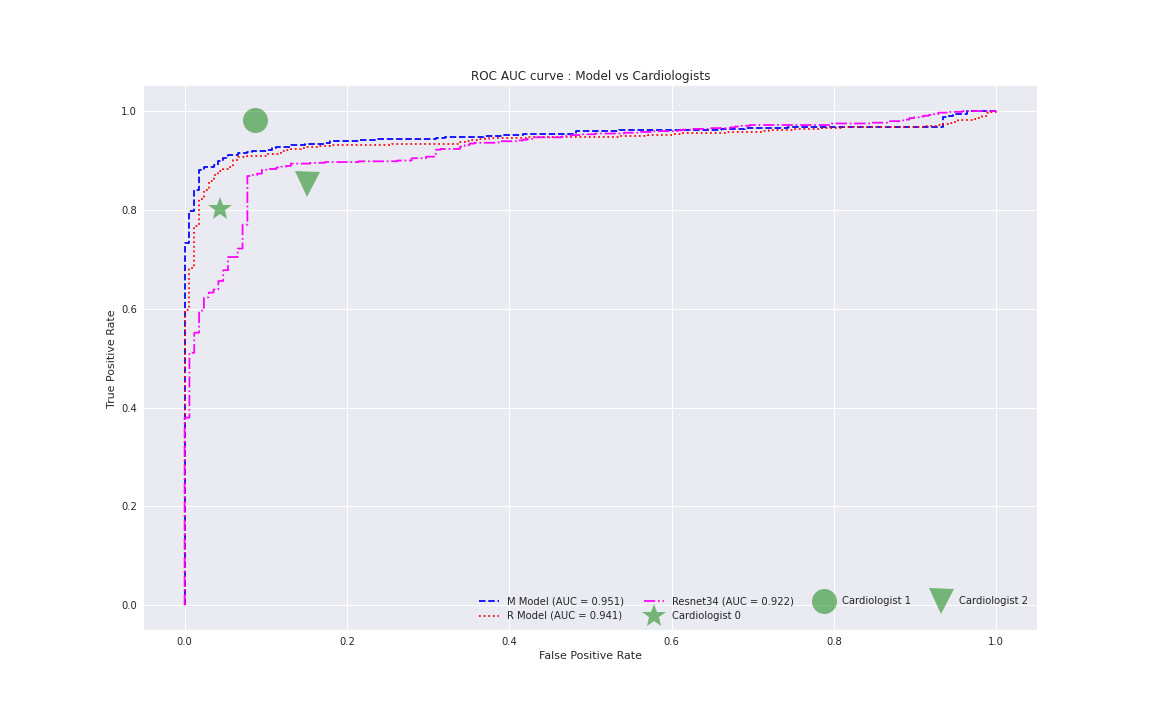}
\centering
\caption{ROC curve : models vs cardiologists}
\label{fig:AUC}
\end{figure}

\begin{figure}[H]
\includegraphics[width=1.0\textwidth]{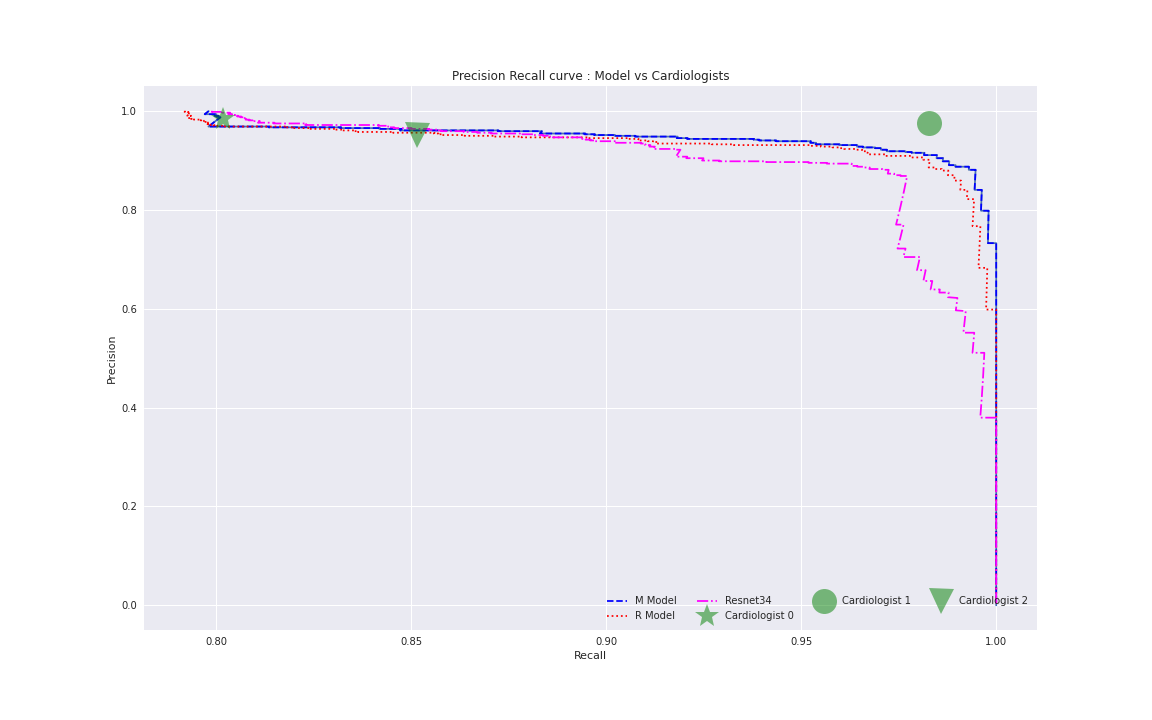}
\centering
\caption{Precision Recall curve : models vs cardiologists}
\label{fig:PR}
\end{figure}

\section{Deployment}
\begin{figure}[h!]
\includegraphics[width=1.0\textwidth]{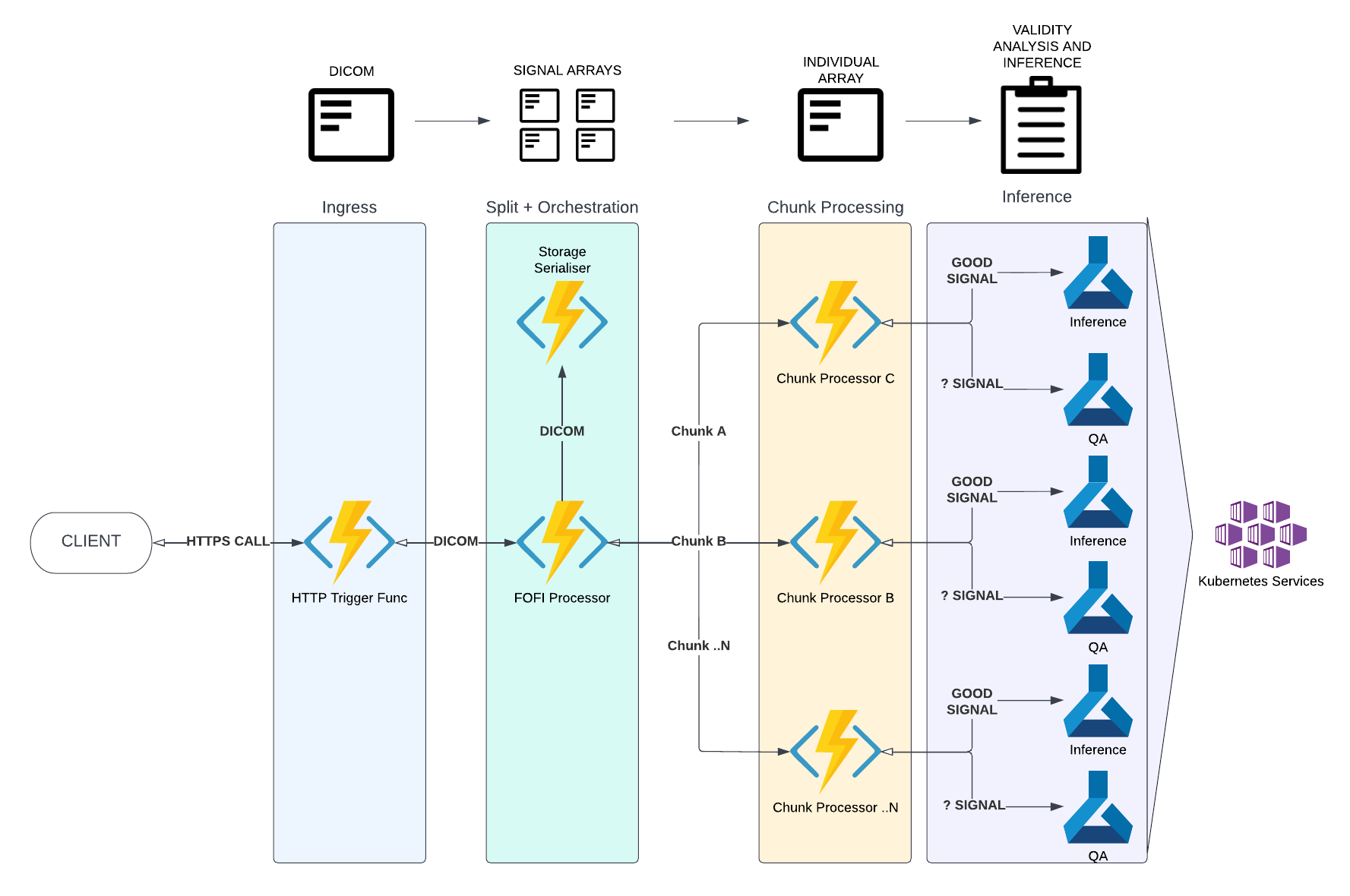}
\centering
\caption{High level functional diagram}
\label{fig:highlevelarch}
\end{figure}

For deployment of the system, we elected to follow a microservices-driven architecture, hosted in the Microsoft Azure Cloud across a collection of Azure services. Summarised, these services were:
\begin{description}
    \item[DevOps] Utilized for Git repositories, pipelines and build agents.
    \item[Function Apps] Utilized as an HTTP middleware to facilitate communication between model endpoints and client applications.
    \item[Machine Learning] Utilized for experimentation, log management, artifact management and as the interface for Azure Kubernetes-hosted model endpoints.
    \item[Kubernetes Service] Utilized as the infrastructure layer for hosting of model endpoints.
    \item[Storage Accounts] Utilized for storage of ingested signal data for the purposes of retraining and analysis.
    \item[Application Insights] Utilized for application monitoring for both the Function App hosted middleware, and the Azure Kubernetes Service hosted model endpoints.
\end{description}

In this instance, inference is performed on a collection of independent 8 second sequences, with individual model responses then aggregated to a reported result over all sequences. As there is no need to communicate state between sequence instances, or for these sequences to be executed consecutively, it can be considered an "embarrassingly parallel" problem. For this problem, we elected to use the "Durable Functions" extension of Azure Function Apps. Durable Functions allows for the creation of more nuanced orchestrations of computation, while maintaining the inherent scalability and fault tolerance of a normal Function App, which are important for the usability of the solution.
\newline
In pseudocode, this orchestration can be seen as:

\begin{mdframed}
\begin{lstlisting}
array_of_signals = transform_input_signal_into_chunk_array(input_signal)

array_of_tasks = [empty array]

for each chunk in array_of_signals
    in_progress_call = call_inference_activity_on_chunk(chunk)
    append in_progress_call to array_of_tasks
    
collected_responses = await array_of_tasks
summarised_response = aggregate_responses(collected_responses)

return summarised_response
\end{lstlisting}
\end{mdframed}

By performing the individual sequence processing within separate activity functions, we can leverage the parallelism and fault tolerance of Durable Function apps. In the event that one of the individual executions fails, that individual activity is restarted, rather than the entire orchestration. 

Each of these individual activity function executions communicate (over HTTPS) with a Kubernetes-backed model endpoint, deployed via the Azure Machine Learning (AML) service. These connections are internally routed through an AML-created load balancer pod layer, to the model endpoints where inference is then performed (both hosted within the Azure Kubernetes Service).

This enables a level of elasticity within the system: model endpoints can be created and destroyed dynamically, and a system "state" can be maintained, where state can be seen as the number of active endpoints (currently performing inference) and readied endpoints (endpoints that are available for inbound connections but idle). This state is maintained via two "levers": internally to Kubernetes, where extra model replica pods are created when there is additional load into the system, and at an infrastructure level, where extra nodes (virtual machines) can be allocated to Kubernetes, thus leading to capacity of:

\begin{equation*}
    available\,endpoints \approx {node\,pool\,count} * (\frac{total\,node\,resources}{single\,instance\,utilisation})
\end{equation*}
As such, we have multiple levers by which we can dynamically adjust the system in response to demand (either dynamically via AKS, or vertically by adjusting node pool specification for higher/lower SKU machines). This level of flexibility is required for multiple reasons. First, because the system is going to be under variable load over a 24 hour period as a result of requests coming directly from cardiologists, and thus would be cost-inefficient if it was unable to flexibly allocate and deallocate resources. Second, because the endpoints are receiving a single short segment of a larger signal, load through to the model endpoints is also dependent on signal length. For instance, a two minute ECG signal, separated into 8 second sequences with 1 second of overlap leads to 17+ separate model endpoint calls (allowing for call failures and retries). Signals are not of uniform length, and so the number of calls will rise and fall dependent on their duration.

\subsection{MLOps and Integration}
"MLOps" refers to a loosely associated set of principles, partially derived from "DevOps", specific to the domain of machine learning systems. As a result, there is no universally-agreed set of features that an "MLOps-guided" system should exhibit, but we have placed importance upon the following principles:
\begin{itemize}
    \item Results of model training runs, model weight versions should be "indefinitely stored" with the ability to deploy and redeploy inference components with newer or older versions of model weights.
    \item Deployment of system components should be automated, without the need for a person to manually deploy or configure individual services.
    \item The system should be "transparent", where any individual component can be monitored with low effort.
    \item Deployment of the system and configuration of the components should ultimately be stored and executed via code, rather than front-end or UI components.
    \item There should be no doubt that the system will operate as expected in the production environment. Conversely, it should be impossible to deploy a version of the system where certain behaviours are not guaranteed.
\end{itemize}

With these principles in mind, the need for a thorough DevOps/MLOps arrangement was made clear. This arrangement was enabled via Azure DevOps (ADO) and the Azure Machine Learning service; primarily orchestrated via Git commit and pull triggers. As the models are deployed on a microservice architecture, we're able to individually update and configure components without redeploying or altering adjacent components (barring API changes). Via ADO build and release pipelines, these changes can also be sequentially deployed through DEV, QA and PROD environments, to reduce the chance of unexpected errors when they arrive in PROD.

\begin{itemize}
    \item \textbf{Merge to main of middleware code.} Triggers a pipeline where middleware changes are sequentially deployed and tested across environments.
    \item \textbf{Merge to main of model training code.} Triggers a pipeline where a new batch AML endpoint for the model training is sequentially deployed, a training run is initiated, and a new model artifact is produced along with error metrics. 
    \item \textbf{New version of the model weights registered to Azure Machine Learning model registry.} Leads to an evaluation of whether the new model is of comparable performance. If they are, the inference webservices are updated with the new model.
    \item \textbf{Merge to main of inference code.} Triggers a pipeline where the new inference code is sequentially deployed and tested across environments.
\end{itemize}

Following these controls, we're able to maintain the system on a service-by-service basis, while being able to effectively manage the complexity of the overall interactions. Monitoring of the individual components was enabled by Azure Application Insights, which integrates with Azure Functions and AML-backed Kubernetes deployments, and yields telemetry such as application logs, response times, and individual node statuses. 

\section{Discussion}
\par 
\subsection{Model Performance}
Our experimental results demonstrate that a deep neural network architecture can render solid results in detection of normal vs abnormal ECG records for canines, and most of the time perform on par with trained cardiologists. Furthermore, we saw that the continuous wavelet approach associated to the augmentation paradigm allows to handle the signal data as an image and then to leverage classical deep learning technics. In particular, it seems that our custom models are performing slightly better comparatively to pre-trained resnet models.
\par Last but not least, the deployment of the solution in the cloud with the activation of MLOps features makes the entire solution more flexible, reliable and scalable.

\subsection{Application and Future Work}
\par Cardiac arrhythmias are associated with an increased risk of sudden death in dogs\cite{Santilli2021-vh}\cite{Borgeat2021-ze}. Cardiac arrhythmias can be found in dogs who are considered healthy, indicating a need for increased screening \cite{Duerr2007-tw}\cite{Silveira2018}. However, diagnosis of canine arrhythmias is a challenge for many veterinarians, due to a combination of factors \cite{hellemans2022diagnosis}. Cited issues include lack of access to ECG monitoring capability or lack of training in ECG interpretation. Automated tools such as the one developed in this work can improve decision support and enhance quality of care available to pets. Normal vs abnormal classifcation provides a valuable first step to automated ECG diagnosis of individual rhythms. Different cardiac arrhythmias and underlying cardiac pathologies require individualized treatment \cite{electrocardiography}. For this reason future work will focus on multi-label diagnosis of ECG abnormalities including arrhtyhmias. Prediction of specific arrhythmias and abnormalities will allow veterinarians to make more complex decisions in real-time including referral to cardiology specialist. 

\section{References}

\bibliographystyle{unsrtnat}
\bibliography{references}

\end{document}